\documentstyle[11pt,epsfig]{article}
\newcounter{subequation}[equation]

\makeatletter

\def\thesubequation{\theequation\@alph\c@subequation}
\def\@subeqnnum{{\rm (\thesubequation)}}
\def\slabel#1{\@bsphack\if@filesw {\let\thepage\relax
   \xdef\@gtempa{\write\@auxout{\string
      \newlabel{#1}{{\thesubequation}{\thepage}}}}}\@gtempa
   \if@nobreak \ifvmode\nobreak\fi\fi\fi\@esphack}
\def\subeqnarray{\stepcounter{equation}
\let\@currentlabel=\theequation\global\c@subequation\@ne
\global\@eqnswtrue
\global\@eqcnt\z@\tabskip\@centering\let\\=\@subeqncr
$$\halign to \displaywidth\bgroup\@eqnsel\hskip\@centering
  $\displaystyle\tabskip\z@{##}$&\global\@eqcnt\@ne
  \hskip 2\arraycolsep \hfil${##}$\hfil
  &\global\@eqcnt\tw@ \hskip 2\arraycolsep
  $\displaystyle\tabskip\z@{##}$\hfil
   \tabskip\@centering&\llap{##}\tabskip\z@\cr}
\def\endsubeqnarray{\@@subeqncr\egroup
                     $$\global\@ignoretrue}
\def\@subeqncr{{\ifnum0=`}\fi\@ifstar{\global\@eqpen\@M
    \@ysubeqncr}{\global\@eqpen\interdisplaylinepenalty \@ysubeqncr}}
\def\@ysubeqncr{\@ifnextchar [{\@xsubeqncr}{\@xsubeqncr[\z@]}}
\def\@xsubeqncr[#1]{\ifnum0=`{\fi}\@@subeqncr
   \noalign{\penalty\@eqpen\vskip\jot\vskip #1\relax}}
\def\@@subeqncr{\let\@tempa\relax
    \ifcase\@eqcnt \def\@tempa{& & &}\or \def\@tempa{& &}
      \else \def\@tempa{&}\fi
     \@tempa \if@eqnsw\@subeqnnum\refstepcounter{subequation}\fi
     \global\@eqnswtrue\global\@eqcnt\z@\cr}
\let\@ssubeqncr=\@subeqncr
\@namedef{subeqnarray*}{\def\@subeqncr{\nonumber\@ssubeqncr}\subeqnarray}
\@namedef{endsubeqnarray*}{\global\advance\c@equation\m@ne%
                           \nonumber\endsubeqnarray}

\makeatletter
\@addtoreset{equation}{section}
\makeatother
\renewcommand{\theequation}{\thesection.\arabic{equation}}

\def\dalemb#1#2{{\vbox{\hrule height .#2pt
        \hbox{\vrule width.#2pt height#1pt \kern#1pt
                \vrule width.#2pt}
        \hrule height.#2pt}}}

\def\half{{\textstyle{1\over2}}}
\let\a=\alpha \let\b=\beta  \let\d=\delta \let\e=\epsilon
  \let\q=\theta  \let\k=\kappa
\let\l=\lambda \let\m=\mu \let\n=\nu  \let\p=\pi \let\r=\rho
\let\s=\sigma \let\t=\tau  \let\f=\phi \let\c=\chi 
       \let\D=\Delta  \let\L=\Lambda
 \let\P=\Pi   \let\F=\Phi 
\let\C=\Chi 
\let\la=\label  
  
\def\nn{\nonumber} \def\bd{\begin{document}} \def\ed{\end{document}}
\def\ds{\documentstyle} \let\fr=\frac \let\bl=\bigl \let\br=\bigr
\let\Br=\Bigr \let\Bl=\Bigl
\let\bm=\bibitem
\let\na=\nabla
\let\pa=\partial \let\ov=\overline
\def\ie{{\it i.e.\ }}
\newcommand{\be}{\begin{equation}}
\newcommand{\ee}{\end{equation}}
\def\ba{\begin{array}}
\def\ea{\end{array}}
\def\ft#1#2{{\textstyle{{\scriptstyle #1}\over {\scriptstyle #2}}}}
\def\fft#1#2{{#1 \over #2}}
\def\del{\partial}
\def\sst#1{{\scriptscriptstyle #1}}
\def\oneone{\rlap 1\mkern4mu{\rm l}}
\def\e7{E_{7(+7)}}
\def\td{\tilde}
\def\wtd{\widetilde}
\def\im{{\rm i}}
\def\bog{Bogomol'nyi\ }
\def\q{{\tilde q}}
\def\hast{{\hat\ast}}
\def\0{{\sst{(0)}}}
\def\1{{\sst{(1)}}}
\def\2{{\sst{(2)}}}
\def\3{{\sst{(3)}}}
\def\4{{\sst{(4)}}}
\def\5{{\sst{(5)}}}
\def\6{{\sst{(6)}}}
\def\7{{\sst{(7)}}}
\def\8{{\sst{(8)}}}
\def\n{{\sst{(n)}}}
\def\oo{{\"o}}
\def\hA{\hat{\cal A}}
\def\ns{{\sst {\rm NS}}}
\def\rr{{\sst {\rm RR}}}
\def\tH{{\widetilde H}}
\def\tB{{\widetilde B}}
\def\cA{{\cal A}}
\def\cF{{\cal F}}
\def\tF{{\wtd F}}
\def\Z{\rlap{\sf Z}\mkern3mu{\sf Z}}
\def\ep{{\epsilon}}
\def\IIA{{\rm IIA}}
\def\IIB{{\rm IIB}}
\def\ads{{\rm AdS}}
\def\R{\rlap{\rm I}\mkern3mu{\rm R}}
\def\mapright#1{\smash{\mathop{-\!\!\!-\!\!\!-\!\!\!-\!\!\!-\!\!\!
             \longrightarrow}\limits^{#1}}}

\def\ba {\begin{eqnarray}}
\def\ea {\end{eqnarray}}
\def\nn {\nonumber}
\def\half{{1\over2}}
\def\a  {\alpha}
\def\b  {\beta}
\def\c  {\gamma}
\def\C  {\Gamma}
\def\d  {\delta}
\def\D  {\Delta}
\def\e  {\epsilon}
\def\F  {\Phi}
\def\k  {\kappa}
\def\l  {\lambda}
\def\L  {\Lambda}
\def\m  {\mu}
\def\n  {\nu}
\def\o  {\omega}
\def\O  {\Omega}
\def\p  {\pi}
\def\P  {\Pi}
\def\r  {\rho}
\def\th {\theta}
\def\s {\sigma}
\def\t  {\tau}
\def\la {\label}
\def\le {\left}
\def\ri {\right}
\def\pa {\partial}
\def\f {\frac}
\def\sq {\sqrt}
\def\no {\noindent}
\def\bi {\begin{itemize}}
\def\ei {\end{itemize}}
\def\np {\newpage}
\def\ra {\rangle}
\def\vs {\vspace}
\def\llra {\Longleftrightarrow}
\def\veck {\vec{k}_\perp}
\def\vecm {|\vec{k}_\perp|}
\def\pl {{\cal P}}
\def\bfy{{\bf y}}
\def\bfk{{\bf k}}

\def\Ei{{\hbox{Ei}}}
\def\Ci{{\hbox{Ci}}}
\def\Si{{\hbox{Si}}}

\newcommand{\ho}[1]{$\, ^{#1}$}
\newcommand{\hoch}[1]{$\, ^{#1}$}
\newcommand{\bea}{\begin{eqnarray}}
\newcommand{\eea}{\end{eqnarray}}
\newcommand{\lra}{\longrightarrow}
\newcommand{\Lra}{\Leftrightarrow}
\newcommand{\aap}{\alpha^\prime}
\newcommand{\bp}{\tilde \beta^\prime}
\newcommand{\tr}{{\rm tr} }
\newcommand{\Tr}{{\rm Tr} }
\newcommand{\NP}{Nucl. Phys. }

\newcommand{\brussels}{\it Physique Th\'eorique et Math\'ematique,
Universit\'e Libre de Bruxelles,\\ Campus Plaine C.P. 231, B-1050
Bruxelles, Belgium}

\newcommand{\unb}{\it Department of Mathematics and Statistics and Department 
of Physics,\\University of New Brunswick, Fredericton, N.B. E3B 5A3 
Canada}

\newcommand{\ucd}{\it Department of Physics,\\
University of California, Davis, CA 95616, USA}

\newcommand{\auth}{J. Gegenberg\hoch{\dagger \sharp 1}, S. Vaidya
\hoch{\ddagger 2} and J.F. V\'{a}zquez-Poritz\hoch{\sharp 3}}

\begin{document}
\begin{flushright}
ULB-TH/02-16\\
UNB-Math 02-02\\
UCD-2002-07 \\
June 2002\\
\hfill{\bf hep-th/0205276}\\
\end{flushright}

\begin{center}

{\large {\bf Thurston Geometries from Eleven Dimensions}}

\vspace{20pt}

\auth

\vspace{10pt}
\hoch{\sharp}\brussels\\
\vspace{10pt}
\hoch{\dagger}\unb\\
\vspace{10pt}
\hoch{\ddagger}\ucd\\

\vspace{30pt}

\underline{ABSTRACT}
\end{center}

In three dimensions, a `master theory' for all Thurston geometries
requires imaginary flux. However, these geometries can be obtained from
physical three-dimensional theories with various additional scalar fields,
which  can be interpreted as moduli in various compactifications of a 
higher-dimensional `master theory'. Three Thurston geometries are of the
form $N_2\times S^1$, where $N_2$ denotes a two-dimensional Riemannian space of 
constant curvature. This enables us to twist these spaces, via T-duality,
into other Thurston geometries as a $U(1)$ bundle over $N_2$. In this way,
Hopf T-duality relates all but one of the geometries in the 
higher-dimensional M-theoretic framework. The exception is the `Sol
geometry,' which results from the dimensional reduction of the decoupling
limit of the D3-brane in a background $B$ field.

{\vfill\leftline{}\vfill\vskip 10pt \footnoterule {\footnotesize
\hoch{1} lenin@math.unb.ca

\hoch{2} vaidya@dirac.ucdavis.edu

\hoch{3} jvazquez@ulb.ac.be

\vskip  -12pt} \vskip   14pt
}

\pagebreak
\setcounter{page}{1}

\section{Introduction}

There is a powerful but under-utilized technology available for dealing
with three-manifolds, due to the work of mathematicians in the 1970's and 
1980's and culminating in Thurston's Geometrization Conjecture 
\cite{thurston}\footnote{A fairly clear exposition of this can be found in the 
review article by P. Scott \cite{scott}.  There has been some use made of
the Geometrization Conjecture in cosmology, beginning with the paper by
Fagundes \cite{fag}.  A non-exhaustive list of other work along these 
lines is in \cite{kodama, yasuno,weeks,piot,koike}.  Examples of attempts to 
understand the proof of the conjecture using techniques from gravitional and 
particle physics are \cite{isen,braham}.  The role of the 
conjecture in high-energy physics is explored in \cite{wael,alvarez}.}.  
This conjecture states that a three-manifold with a given 
{\it topology} has a canonical decomposition into a connected sum of `simple 
three-manifolds,' each of which admits one, and only one, of eight
homogeneous {\it geometries}: $H^3$, $S^3$, $E^3$, $S^2\times S^1$, 
$H^2\times S^1$, Sol, Nil and SL(2,R). The conjecture has not been
completely proven but few in the field doubt its veracity.  

To see why this technology would be important in fundamental physics,
consider the case of two-dimensional physics, such as conformal field
theories or the world-volume of string theory.  In the path-integral
formalism, one must sum over two-dimensional topologies and
geometries.  This procedure is executable for two-dimensional Riemannian
manifolds because of the existence of a geometrization {\it theorem} 
 \cite{poincare}.  Namely, a closed two-dimensional manifold with handle
number (i.e. topology) $h=0$, $1$ or $\geq 2$, respectively, admits 
the spherical, torus, or hyperbolic geometry.   Hence, in the path 
integral, we can sum over deformations of each of these geometries, then sum 
over the handle number.  

In (super)membrane physics \cite{mem} or in three-dimensional quantum gravity
we should be able to perform path-integral quantization via a similar 
procedure to that in two dimensions.  However, it is crucial to understand
the connection between three-dimensional topology and M-theory in an
analogous manner to the two-dimensional case, where two-dimensional topology 
provides the stringy analogue of Feynman diagrams. We are a long way from
this but as a first step we search for a theory which admits the eight
Thurston spaces as solutions.  

The plan of this paper is as follows. In section 2, we discuss the
difficulties involved in attempting to obtain all eight Thurston spaces
from a single `master theory' in three dimensions. This serves as 
motivation to search for a `master theory' in higher dimensions. In
section 3, we show how three of the Thurston geometries are of the form 
$N_2\times S^1$, where $N_2$ denotes $H^2$, $E^2$ or $S^2$. In an M-theoretic 
context, these geometries can be T-dualized into other Thurston geometries
which are a $U(1)$ bundle over $N_2$. The only Thurston space which does
not naturally fit into this scheme is Sol, which results from the
dimensional reduction of the decoupling limit of the D3-brane in a
background $B$ field. We present conclusions in section 4.

\section{Difficulties finding `master theory' in three
dimensions}

One can construct three-dimensional Chern-Simons gauge theories in which each
gauge group is the group of isometries of a given Thurston geometry.  This
is well-known for the isotropic geometries $E^3$, $S^3$ and $H^3$, and can
also be done for the anisotropic Thurston geometries\footnote{For example, 
the isometry group of Sol is the solvable Lie Group with the three generators
${T,L,M}$ which satisfy the algebra $[T,L]=-L;[T,M]=M;[L,M]=0$.
The gauge potential is ${\cal A}={\cal A}^1T+{\cal A}^2M+{\cal A}^3L$. The
Chern-Simons equation of motion is $F({\cal A})=d{\cal A}+{1\over2}[{\cal
A},{\cal A}]=0$, which has a solution ${\cal A}^1=dx;\\
{\cal A}^2=e^{-x}dz; {\cal A}^3=e^xdy$. These are the frame-fields for
metric corresponding to Sol.}. Of the Thurston spaces, only $E^3$, $S^3$
and $H^3$ are solutions of Einstein gravity. In search of a single theory 
from which all eight of the Thurston geometries arise, we next turn to the 
low-energy limit of three-dimensional string theory, which has a metric 
$g_{\mu\nu}$, dilaton $\phi$, Abelian 2-form potential $B_{(2)}$ with
field strength $H_{(3)}=dB_{(2)}$ and a `constant' term in the level of
the original sigma model \cite{3dstring,kal}. Finding that this theory
fares no better than pure gravity as a `master theory,' we consider a
hypothetical three-dimensional theory which is identical to the low-energy
limit of three-dimensional string theory, except for an additional Abelian
gauge field $A_{(1)}$. This hypothetical theory will serve as a straw man
to be knocked down, in order to effectively demonstrate the difficulties
encountered in our search for a three-dimensional `master theory'. The
corresponding action is given by
\be 
S=\int d^3x\sqrt{g}\, e^{-2\phi}\left({4\over k}+R+4|\nabla\phi|^2 
-{1\over12}H_{\mu \nu \rho}H^{\mu \nu \rho}-{1\over2}
F_{\mu \nu}F^{\mu \nu} \right)+{e\over3} 
\epsilon^{\mu\nu\rho}A_{\mu} F_{\nu \rho}, \label{3daction}
\ee
where the last term is the Abelian Chern-Simons term for the field
$A_{(1)}$, and $F_{(2)}=dA_{(1)}$. All the Thurston geometries are solutions 
of the equations of motion of this theory for various values of the level
$k$ and the constant $e$, as well as the other fields. 

With the exceptions of Sol and $H^3$, the Thurston spaces can be 
characterized topologically as Seifert fibre bundles $\eta$ over an orbifold 
$Y$.  The topology of a Seifert fibre bundle is determined by the Euler number 
$\chi(Y)$ of $Y$ and the Euler number $e(\eta)$ of the bundle $\eta$ 
\cite{thurston,scott,alvarez}.  It turns out that the level $k$ is related 
to $\chi(Y)$ by $\chi(Y)=4/k$ for all the Thurston spaces except Sol and 
$H^3$.  In addition, the constant $e$ turns out to be precisely
the Euler number $e(\eta)$.

However, this is a `master theory' only formally since not all of the
Thurston geometries arise as {\it physical} solutions. The three-form field 
strength $H_{(3)}$ is real only for the Thurston spaces which are Seifert 
fibre bundles. Specifically, Sol and $H^3$ require an imaginary $H_{(3)}$.
In addition, $H^2\times S^1$, Nil and SL(2,R) require an imaginary
$F_{(2)}$.

In order for all Thurston solutions to have real fields in a hypothetical 
three-dimensional theory of sufficient generality, we could
embark on the guessing game of adding, for example, more scalar
fields. For instance, Sol arises as a physical solution when there are
three scalars within the theory. However, at the end of this laborious
procedure, one is left with a model that is not well-motivated from
the three-dimensional viewpoint. We have arbitrarily added fields and
do not even know the precise way in which they interact with the
already-present fields. However, one could quite naturally regard the
additional scalars as being the moduli of compact spaces, due to
various reductions from a `master theory' in higher dimensions. Also,
the curvatures of the compact spaces yield effective cosmological
terms.

\section{Thurston geometries from eleven dimensions}

\subsection{Twisted spaces}

Various $d$-dimensional geometries $M_d$ can be expressed as $U(1)$
bundle over $(d-1)$-dimensional geometries $N_{d-1}$. We will consider the case
of $d=3$, for which the metric of the twisted space has the form
\be
dM_3^2=dN_2^2+(dz+A_{(1)})^2,
\ee
with $N_{(2)}=dA_{(1)}$ the volume form on $N_2$. 

There are three compact, locally homogeneous Riemannian spaces in two
dimensions, which we label as $N_2$ in order to represent $S^2$, $H^2$
or $T^2$. In three dimensions, according to Thurston's Geometrization
Conjecture \cite{thurston}, there are eight of such spaces. A direct product 
of $N_2\times S^1$ produces three of the Thurston spaces. A $U(1)$ bundle
along $S^1$ and around $N_2$ twists these previous direct product spaces
into four other Thurston spaces \cite{thurston, scott, alvarez}.

To be explicit, $S^3$ as a $U(1)$ bundle over $S^2$, is expressed as
\be
ds_3^2=\ft14 \Big( d\theta^2+\sin^2 \theta\, d\phi^2+(d\psi+\cos \theta\,
d\phi)^2\Big).
\ee
Nil, as a bundle over $T^2$, is given by
\be
ds_3^2=dx^2+dy^2+(dz-x\, dy)^2.
\ee
$H^3$ can be expressed as twisted $H^2\times S^1$ only in the unphysical
case of an imaginary $U(1)$ field:
\be
ds_3^2=\ft14 \Big( {\rm cosh}^2 \rho\, dt^2+d\rho^2+(dz+i\, {\rm sinh}
\rho\, dt)^2 \Big). \label{H3}
\ee
On the other hand, SL(2,R) can be expressed as twisted $H^2\times S^1$,
as in the above equation, except with a real $U(1)$ field\footnote{$H^3$
expressed as an imaginary bundle over $H^2$ can be obtained as a Wick
rotation of $AdS_3$ expressed as a bundle over $AdS_2$. The alternative
Euclideanization of $g_{tt}\longrightarrow -g_{tt}$ produces SL(2,R) as
a real bundle over $H^2$. However, since Wick rotations do not ensure that
Majorana or self-duality constraints are compatible with the Euclidean
signature, it is best to begin with a solution which contains a
(Riemannian) hyperbolic factor from the start \cite{cremmer}.}. This means 
that T-duality can physically relate the geometry $H^2\times S^1$ with 
SL(2,R) but not with $H^3$. Thus, in particular, from a supergravity
geometry which contains $T^2\times S^1$ or $H^2\times S^1$, one can
T-dualize to obtain a solution which contains the less common geometries
Nil or SL(2,R), respectively.

Direct products of hyperbolic, spherical and Euclidean spaces frequently
result as vacua in the near-horizon regions of $p$-brane solutions in
string/M-theory. As an example, the near-horizon region geometry of a
standard M2/M2/M5/M5 brane intersection can be expressed as 
$(AdS_2 \times S^1)\times (S^2\times S^1)\times (T^2\times S^1)\times T^2$. 
With the appropriate $D=11$ three-form field\footnote{Hopf T-duality
untwists Thurston spaces into the form $N_2\times S^1$ only when there are 
only R-R charges present in the ten-dimensional theory\cite{untwist2}.}, 
T-duality serves to twist the above product spaces in the parentheses, in
order to obtain other Thurston spaces embedded within the M-theoretic
context. T-duality along a $U(1)$ fibre is known as Hopf T-duality
\cite{untwist1, untwist2}. The connection between the twisted spaces and
direct product spaces is summarized in Table 1.

\bigskip
\begin{center}
\begin{tabular}{|c|c|c|}\hline
Product Space ($N_2\times S^1$) & Hopf T-duality & Twisted Space ($M_3$)\\
\hline\hline
$S^2\times S^1$ & $\longleftrightarrow$ & $S^3$ \\
$T^2\times S^1$ & $\longleftrightarrow$ & Nil \\
$H^2\times S^1$ & $\longleftrightarrow$ & $H^3$,\ SL(2,R) 
\\ \hline
\end{tabular}
\end{center}

\centerline{Table 1: Twisted Thurston spaces}
\bigskip

It is interesting to note that the Thurston spaces which are related to
each other by Hopf T-duality share the same value for the level $k$ in the
hypothetical three-dimensional theory of Section 2.  Only Sol does not 
appear in Table 1.

\subsection{Sol from D3-brane in $B$ field}

Consider the decoupling limit of a D3-brane with a constant NS $B$ field,
for which the $B$ field goes to infinity. The metric solution is given by
\cite{hash,russo}
\be
ds_{10}^2=\Big( \frac{u}{R}\Big)^2 \Big( -dt^2+dx_1^2+\frac{1}{1+b^2
(u/R)^4}(dx_2^2+ x_3^2)\Big) +\Big( \frac{R}{u}\Big)^2 du^2+R^2
d\Omega_5^2,\label{solmetric5}
\ee
where $R$ corresponds to the D3-brane charge and $b$ is proportional to
the $B$ field, which we take to both be unity for simplicity. 
This is the metric of the supergravity dual to a Yang-Mills
gauge theory with noncommuting $x_2,x_3$ coordinates. For small $u$, the
above geometry reduces to $AdS_5 \times S^5$, corresponding to ordinary
Yang-Mills occupying the IR region of the dual field theory.
Using the metric Ansatz \cite{reduce}
\be
ds_D^2={\rm e}^{-2\alpha \varphi}ds_d^2+{\rm e}^{2(d-2)\alpha \varphi}
dz^2,
\ee
where $\alpha=-1/\sqrt{2(d-2)(D-2)}$, we dimensionally reduce on $S^5$,
$t$ and $x_3$. Approaching the boundary at $u=\infty$ yields the metric
for Sol, given by
\be
ds_3^2=u^2 dx_1^2+\frac{dx_2^2}{u^2}+\frac{du^2}{u^2}.
\ee
Note that all the fields in our three-dimensional hypothetical model of
Section 2, for which Sol is a solution, are accounted for via the
above dimensional reduction from ten dimensions. In particular, the
five-form field strength and $B$ field dimensionally reduce to three and
two-form field strengths respectively. Also, the ten-dimensional dilaton
and moduli of $t$ and $x_3$ correspond to the three scalar fields in our
hypothetical model. 

Since we took the large $u$ limit in (\ref{solmetric5}), Sol originates
from the dual gravity description of the UV regime of non-commutative 
Yang-Mills.

\section{Conclusions}

Thurston spaces can be obtained as solutions to various Chern-Simons gauge 
theories. The Euclidean path-integral approach to three-dimensional quantum 
gravity and (super) membrane physics motivated us to search for a
single `master theory' for all of the Thurston spaces. In this vein, we
constructed a model theory by adding a gauge field to low-energy string
theory in three dimensions. We found that some of the Thurston solutions
required an imaginary flux unless there are additional scalar fields in
the theory. 

Although such a three-dimensional theory does not give a complete picture
of the Thurston geometries, it yields some interesting hints of the
underlying structure.  This is contained in the correspondence 
between the orbifold topology of six of the eight geometries that are 
Seifert fibre bundles and the parameters $e$ and $k$ of the string 
theory/sigma model whose low-energy effective action is (\ref{3daction}).  It 
would also appear that the fact that Sol and $H^3$ do not admit a Seifert
fibre bundle structure \cite{thurston,scott, alvarez} is related to the 
non-existence of a {\it real} 2-form potential, unless there are additional 
scalar fields.  

Additional scalars arise naturally as the moduli of compact 
dimensions. Also, various cosmological terms required by most of the
Thurston solutions could be considered as the curvatures of compact
spaces. We were thus motivated to look to higher dimensions for
a `master theory' for the Thurston solutions.  Two and three-dimensional
Euclidean, spherical and hyperbolic spaces frequently occur within
M-theoretic vacuum solutions.  T-dualizing such spaces yields additional
Thurston geometries expressed as a $U(1)$ bundle over a two-dimensional
Einstein space. In particular, T-duality twists $T^2\times S^1$ and
$H^2\times S^1$ into Nil and SL(2,R) respectively. The only Thurston space
which does not fit into this scheme is Sol, which is derivable from the
decoupling limit of a D3-brane in a background $B$ field.

Sol originates from the dimensional reduction of a dual gravity description 
of the UV regime of non-commutative Yang-Mills. Whether this indicates that 
Sol may be a useful toy model for the study of holography is still in
the realm of speculation. The scalar wave equation on the Sol background,
as for the ten-dimensional origin, is a Mathieu equation. This would
indicate that the correlation functions in our toy model would have the
same form as that derived from ten-dimensional theory. Sol has boundaries
at $u=0$ and $u=\infty$, and holographical dual field theory would reside
at the latter boundary.

Perhaps understanding how the Thurston spaces arise from the deep well of
M-theory may indicate whether the Euclidean path integral approach to
quantum gravity lies within a corner of M-theory.  We are currently
exploring issues involving the topology of the Thurston spaces and the 
structure of the fields in the three and higher-dimensional theories of 
origin.

\section*{Acknowledgments}

J.G. is partially supported by the Natural Sciences and Engineering Research 
Council of Canada. S.V. is partially supported by US Department of Energy
grant DE-FG03-91ER40674. J.F.V.P. is supported in full by the Francqui
Foundation (Belgium), the Actions de Recherche Concert{\'e}es of the
Direction de la Recherche Scientifique - Communaut\'e Francaise de
Belgique, IISN-Belgium (convention 4.4505.86).

\end{document}